\documentclass[preprint,authoryear]{elsarticle}

\usepackage{lineno}
\usepackage[colorlinks]{hyperref}
\usepackage{graphicx}
\usepackage{amsmath}
\usepackage{hhline}
\usepackage{makecell}
\usepackage{booktabs}   
\usepackage{multirow}
\usepackage{xcolor}
\usepackage{xurl}
\usepackage{subcaption}
\usepackage{hyperref}
\usepackage{float}
\usepackage{threeparttable}
\usepackage{footmisc}
\newcolumntype{B}{>{\centering\arraybackslash}m{3cm}}
\newcolumntype{L}{>{\centering\arraybackslash}m{2.5cm}}

\DeclareMathOperator*{\argmax}{arg\,max}
\DeclareMathOperator*{\argmin}{arg\,min}

\modulolinenumbers[5]

\journal{Medical Image Analysis}

\biboptions{authoryear}
\begin{document}

\begin{frontmatter}

\title{Adaptation to CT Reconstruction Kernels by Enforcing Cross-domain Feature Maps Consistency}
% \tnotetext[mytitlenote]{Fully documented templates are available in the elsarticle package on \href{http://www.ctan.org/tex-archive/macros/latex/contrib/elsarticle}{CTAN}.}

% %% Group authors per affiliation:
% \author{Elsevier\fnref{myfootnote qwerty}}
% \address{Radarweg 29, Amsterdam}
% \fntext[myfootnote]{Since 1880.}

\author[mipt]{Stanislav Shimovolos}
\author[mipt]{Andrey Shushko}
\author[skoltech,airi]{Mikhail Belyaev}
\author[skoltech,airi]{Boris Shirokikh\corref{mycorrespondingauthor}}
\cortext[mycorrespondingauthor]{Corresponding author}
\ead{boris.shirokikh@skoltech.ru}

\address[mipt]{Moscow Institute of Physics and Technology, Moscow, Russia}
\address[skoltech]{Skolkovo Institute of Science and Technology, Moscow, Russia}
\address[airi]{Artificial Intelligence Research Institute (AIRI), Moscow, Russia}
% \address[iitp]{Kharkevich Institute for Information Transmission Problems, Moscow, Russia}

\begin{abstract}
Deep learning methods provide significant assistance in analyzing coronavirus disease (COVID-19) in chest computed tomography (CT) images, including identification, severity assessment, and segmentation. Although the earlier developed methods address the lack of data and specific annotations, the current goal is to build a robust algorithm for clinical use, having a larger pool of available data. With the larger datasets, the domain shift problem arises, affecting the performance of methods on the unseen data. One of the critical sources of domain shift in CT images is the difference in reconstruction kernels used to generate images from the raw data (sinograms). In this paper, we show a decrease in the COVID-19 segmentation quality of the model trained on the smooth and tested on the sharp reconstruction kernels. Furthermore, we compare several domain adaptation approaches to tackle the problem, such as task-specific augmentation and unsupervised adversarial learning. Finally, we propose the unsupervised adaptation method, called F-Consistency, that outperforms the previous approaches. Our method exploits a set of unlabeled CT image pairs which differ only in reconstruction kernels within every pair. It enforces the similarity of the network’s hidden representations (feature maps) by minimizing mean squared error (MSE) between paired feature maps. We show our method achieving 0.64 Dice Score on the test dataset with unseen sharp kernels, compared to the 0.56 Dice Score of the baseline model. Moreover, F-Consistency scores 0.80 Dice Score between predictions on the paired images, which almost doubles the baseline score of 0.46 and surpasses the other methods. We also show F-Consistency to better generalize on the unseen kernels and without the specific semantic content, e.g., presence of the COVID-19 lesions.

\end{abstract}

% alphabetically
\end{frontmatter}

% \linenumbers

\section{Introduction}
\label{sec:intro}

After the coronavirus disease (COVID-19) outbreak, a wide spectrum of automated algorithms have been developed to provide an assistance in clinical analysis of the virus \citep{shoeibi2020automated}. Among others, we consider the analysis of the chest computer tomography (CT) images. Firstly, CT imaging de facto has become one of the reliable clinical pretests for COVID-19 diagnosis \citep{rubin2020role}. Secondly, well-developed deep learning techniques for volumetric CT processing allow precise and efficient analysis of the different COVID-19 markers. The latter includes identification \citep{song2021deep}, prognosis \citep{meng2020deep}, severity assessment \citep{lassau2021integrating}, and detection or segmentation of the consolidation or ground-glass opacity.

One of the easiest to interpret and clinically useful markers is segmentation \citep{shi2020review}. Segmentation provides us with classification, severity estimation \citep{shan2020lung}, or differentiating from other pathologies in a straightforward manner, by evaluating the output mask. However, training a segmentation model takes huge efforts in terms of voxel-wise annotations. The earlier developed models have faced the lack of publicly available data annotated with segmentation masks. To achieve the high segmentation quality, the more sophisticated methods have been designed, e.g., solving a multitask problem, merging datasets with different annotations \citep{goncharov2021ct}. Now, a larger pool of COVID-19 segmentation datasets is available, e.g., \citep{tsai2021rsna}; therefore, the current goal is to build a robust algorithm for clinical use.

In merging a larger pool of data, the problem of \textit{domain shift} arises. Domain shift is one of the most salient problems in medical computer vision \citep{choudhary2020advancing}. A model trained on the data from one distribution might yield poor results on the data from the different distribution. In CT imaging, one of the main sources of the domain shift is the difference in \textit{reconstruction kernels}, the parameter of the Filtered Back Projection (FBP) reconstruction algorithm \citep{schofield2020image}. One could perceptually compare the same image reconstructed with two different kernels in Fig. \ref{fig:consistency_preds}, e.g., B1 and B2. For the kernel-caused domain shift, several works have shown the deterioration of the models' quality, in lung cancer segmentation \citep{ct_conversion_2}, and in emphysema segmentation \citep{ct_conversion_1}.

In this paper, we show that the domain shift induced by the difference in reconstruction kernels decreases the quality of the COVID-19 segmentation algorithms. To do so, we construct two domains from the publicly available data: the \textit{source} domain with the \textit{smooth} reconstruction kernels and \textit{target} domain with the \textit{sharp} reconstruction kernels. See the detailed data description in Sec. \ref{sec:data}. We train the segmentation model on the \textit{source} domain and test it on the \textit{target} domain. With the observed decrease of test score, we then validate the most relevant domain adaptation methods. In our comparison, we include an augmentation approach \citep{saparov2021zero}, unsupervised adversarial learning \citep{uda}, and our proposed feature maps consistency regularization. We describe all methods in Sec. \ref{sec:method}.

Although the augmentation can work only with the \textit{source} data, the adversarial and consistency-based methods require additional unlabeled data from the \textit{target} domain. In this task, a large pool of unlabeled chest CT image pairs which differ only in reconstruction kernels within every pair is publicly available, e.g., \citep{cancer500}. The intuition here is that the adaptation methods should outperform the augmentation one when a broader range of real-world data is available. Furthermore, we propose enforcing the cross-domain feature maps consistency between paired images; we call our method \textit{F-Consistency}. One could find the schematic representation in Fig. \ref{fig:method_schematic}. \textit{F-Consistency} minimizes the mean squared error (MSE) between the network’s hidden representations (feature maps) of paired images. We expect that explicitly enforcing consistency on the paired images should outperform the adversarial learning that emulates the similar behaviour minimizing the adversarial loss.

%  that outperforms the previous approaches.
% we also requier methods to have consistent predictions

We also note that our method could be scaled on the other tasks, such as classification, detection, or multitask, without any restrictions. Below, we discuss the most relevant works to our method, then summarize the contributions.

\subsection{Related work}
\label{ssec:intro:related}

We begin with discussing the task-specific augmentation approaches since they are a straightforward solution to the domain shift problem. Contrary to the classical augmentation techniques for CT images like windowing \citep{windowing} or filtering \citep{filtering}, the authors of \citep{saparov2021zero} proposed FBPAug, augmentation that directly approximates our domain shift. The authors also showed that FBPAug outperforms other augmentations. However, this method has several drawbacks which we discuss in Sec. \ref{ssec:method:fbpaug}. We consider FBPAug as one of the solutions and compare it with the other methods.

With the unlabeled \textit{target} data, we can apply unsupervised domain adaptation methods to improve the model's performance on the \textit{target} domain. The adversarial approaches are shown to outperform other methods \citep{uda}. Considering the paired nature of our data, we divide the adversarial methods into two groups: (i) image-to-image translation and (ii) feature-level adaptation.

The first group of methods aims to translate an image from \textit{source} to \textit{target} domain. In \citep{cycle_gan}, authors used CycleGAN for unpaired image-to-image translation. Further, this method was implemented both for the MRI \citep{cycle_gan_vendors} and CT \citep{cycle_gan_contrast} images translation.  The paired image-to-image translation is closer to our setup. Such a method requires the image pairs that have different style but the same semantic content. CT images that differs only in reconstruction kernel correspond to this setup. Here, the authors of \citep{ct_conversion_1} and \citep{ct_conversion_2} proposed a convolutional neural network (CNN) to translate the images reconstructed with one kernel to another kernel. Despite the reasonable performance, image translation methods lack the generalization ability. If the data consists of more than two domains, we need to train a separate model for every pair. They also do not address the cases with the unseen domains. Therefore, we leave the image translation approaches without consideration.

The other group of methods is independent from the number of domains. Mostly, the feature-level adaptation methods are based on the adversarial learning as in \citep{uda}. The latter approach also finds several successful applications in medical imaging, e.g., \citep{kamnitsas2017unsupervised}, \citep{dou2018unsupervised}. Since these methods are conceptually close to each other, we stick with implementing a deep adversarial neural network, \textit{DANN}, from \citep{uda}.

The idea of F-Consistency is conceptually close to the self-supervised learning, where the unlabeled data is used to pretrain a model using pseudo-labels. In \citep{taleb20203d}, the authors described different pretext tasks in medical imaging. Similarly to our approach, the authors of \citep{constr_learnining} enforce the model consistency for the initial and augmented images at the prediction and feature levels. Furthermore, the authors of \citep{melas2021pixmatch} extended the self-supervised methodology to solve a domain adaptation problem. However, the goal of the self-supervised methods is using a large collection of images without annotations to improve the model's performance on the \textit{source} domain. Contrary to self-supervised learning, our goal is to achieve the highest possible performance on the \textit{target} domain.

Finally, we note that there is no standardized benchmark for the COVID-19 segmentation task \citep{roberts2021common}. We also solve the isolated problem of domain adaptation that could be extended on the classification or detection tasks. Therefore, the comparison with the other COVID-19 segmentation approaches we leave out-of-scope.

% It worth to mention that we do not compare quality of the proposed approach to other COVID-19 segmentation algorithms. The proposed method might be applied to any other machine learning task that is sensitive to the domain shift caused by reconstruction kernels, therefore we compare its quality only to the baseline model.

% Some works extend this methodology to domain adaptation problem. For instance, authors of  use pixel-wise prediction consistency between original target image and pertrubated one to solve domain adaptation problem.

% \todo{other covid algos}

\subsection{Contributions}
\label{ssec:intro:contrib}

Our work highlights a domain shift problem in the COVID-19 segmentation task and suggests an efficient solution to this problem. We summarize our three main contributions as follows:

\begin{itemize}
    \item Firstly, we demonstrate that the difference in CT reconstruction kernels affects the segmentation quality of COVID-19 lesions. The model without adaptation achieves only $0.56$ Dice Score on the unseen domain, while the best adaptation methods scores $0.64$. In terms of similarity between predictions on the paired images, the baseline Dice Score is $0.46$, which is almost two times lower than $0.80$ achieved by our method.
    \item Secondly, we adopt a series of adaptation approaches to solve the highlighted problem and extensively compare their performance under the different conditions.
    \item Thirdly, we propose the flexible adaptation approach that outperforms the other considered methods. We also show our method to better generalize to unseen CT reconstruction kernels and it is less sensitive to the semantic content (COVID-19 lesions) in the unlabeled data.
\end{itemize}

\section{Method}
\label{sec:method}

%\begin{figure}[h!]
%    \centering
%      \caption{Paired image example}%
%    \subfloat[Soft kernel]{{\includegraphics[width=5cm]{figures/scan_a.png} }}
%    \qquad
%    \subfloat[Edge preserving kernel]{{\includegraphics[width=5cm]{figures/scan_b.png} }}
%\end{figure}

In this paper, we consider solving a binary segmentation task, where the positive class is the voxels of volumetric chest CT image with the consolidation or ground-glass opacity. All methods are built upon the convolutional neural network, which we detail in Sec. \ref{ssec:method:covid_segm}.

We train these methods using the annotated dataset $S_s = \{ ( x_i, y_i ) \}_{i=1}^{N_s}$, where $x$ is a volumetric CT image, $y$ is a corresponding binary mask, and $N_s$ is the total size of training dataset. The dataset $S_s$ consists of images reconstructed with \textit{smooth} kernels, and we call it \textit{source} dataset. We test all methods using the annotated dataset $S_t = \{ ( x_i, y_i ) \}_{i=1}^{N_t}$. The testing dataset $S_t$ consists of the $N_t$ images reconstructed with the \textit{sharp} kernels, and we call it \textit{target} dataset. Although $S_t$ contains annotations, we use them only to calculate the final score.

In Sec. \ref{ssec:method:fbpaug}, we describe the only adaptation method, FBPAug, that uses no data except the \textit{source} dataset. The other methods use additional paired dataset $S_2 = \{ ( x_i, \tilde{x}_i ) \}_{i=1}^{N_2}$, which has no annotations. However, every image $x \in S_2$ has a paired image $\tilde{x}$ reconstructed from the same sinogram but with different kernel. Here, we assume that $x$ belongs to the \textit{source} domain and $\tilde{x}$ belongs to the \textit{target} domain. Now, the problem can be formulated as unsupervised domain adaptation, and we detail the corresponding adversarial training approach in Sec. \ref{ssec:method:dann}. We also propose to explicitly enforce the similarity between feature maps of paired images; see Sec. \ref{ssec:method:f-consistency}. In Sec. \ref{ssec:method:p-consistency}, we detail enforcing the similarity between predictions.

\subsection{COVID-19 segmentation}
\label{ssec:method:covid_segm}

In all COVID-19 segmentation experiments, we use the same 2D U-Net architecture \citep{unet} trained on the axial slices. We do not use a 3D model for two reasons. Firstly, as we show in Sec. \ref{ssec:data:segm}, the images have a large difference in the inter-slice distances (from $0.6$ to $8.0$ mm), which can affect the performance of the 3D model. Secondly, authors of \citep{goncharov2021ct} have shown 2D and 3D models yielding similar results in the same setup with various inter-slice distances. Moreover, we note that all considered methods are independent of the architecture choice. We also introduce the standard architectural modifications, replacing every convolution layer with the Residual Block \citep{resnet}. To train the segmentation model, we use binary cross-entropy (BCE) loss. Other training details are given in Sec. \ref{ssec:exp:covid_segm}.

\subsection{Filtered Backprojection Augmentation}
\label{ssec:method:fbpaug}

The first adaptation method that we consider is a task-specific augmentation, called FBPAug \citep{saparov2021zero}. FBPAug emulates the CT reconstruction process with different kernels; thus, it might be a straightforward solution to the domain shift problem, caused by the difference in kernels.

However, FBPAug gives us only an approximate solution, which is also restricted by choice of kernels parameterization. We describe FBPAug as a three-step procedure for a given image $x \in S_s$. Firstly, it applies a discrete Radon transform to the image. Secondly, it convolves the transformed image with the reconstruction kernel. The kernel is randomly sampled from the predefined parametric family of kernels on every iteration. Thirdly, it applies the back-projection operation to the result and outputs the augmented image $FBP(x)$. A complete description of the method could be found in \citep{saparov2021zero}.

We outline two weak spots in the FBPAug pipeline that motivate us to use the other domain adaptation approaches. As described above, FBPAug applies two discrete approximations, a discrete Radon transform, and back-projection, leading to information loss. Furthermore, the original convolution kernels used by CT manufacturers are unavailable, and the parametric family of kernels proposed in \citep{saparov2021zero} is also an approximation. Thus, we expect FBPAug to perform worse than the other adaptation methods when a wider range of paired data is available for the latter methods.

Nevertheless, FBPAug improves the consistency scores in \citep{saparov2021zero}, and we consider it as one of the main adaptation approaches. We give the experimental details in Sec. \ref{ssec:exp:fbpaug}.

\subsection{Deep Adversarial Neural Network}
\label{ssec:method:dann}

Further, we detail the methods that work with the unlabeled pool of (paired) data $S_2$. As mentioned at the beginning of the section, the problem can be reformulated as unsupervised domain adaptation. The most successful approaches to this problem are based on adversarial training. Therefore, we adopt the approach of \citep{uda} and build a \textit{deep adversarial neural network} (DANN). The reason to choose this method we discuss in Sec. \ref{ssec:intro:related}.

DANN includes an additional \textit{domain classifier} or \textit{discriminator} which aims to classify images between the \textit{source} and \textit{target} domains using their feature maps. We train the model to minimize the loss on the primary task (segmentation) and simultaneously maximize the discriminator's loss. Thus, the segmentation part of the model learns domain features that are indistinguishable for the discriminator. The latter should improve the performance of the model on the \textit{target} domain. % In our setup, these domains correspond to images with smooth and sharp reconstruction kernels, respectively.

For the segmentation task, we modify the architectural design of DANN; see Fig. \ref{fig:method_schematic} (6). It consists of three parts: (i) feature extractor $H_f$, the part of segmentation model that maps input images $x$ into the feature space; (ii) segmentation head $H_p$, the complement part of segmentation model that predicts binary mask $\hat{y} = H_p \left( H_f \left( x \right) \right)$; and (iii) discriminator $H_d$, the separate neural network that predicts domain label $\hat{d} = H_d \left( H_f \left( x \right) \right)$. In Fig. \ref{fig:method_schematic}, $H_f$ and $H_p$ correspond to the encoder and decoder parts of the model, respectively. $H_d$ is denoted with the dashed green arrow that passes the aggregated features to the adversarial loss.

Following \citep{uda}, our optimization target is
\begin{align}
    \label{eq:dann_opt}
    \begin{split}
    E(\theta_f, \theta_p, \theta_d) &= \sum_{i=1}^{N_s}L_s(H_p(H_f (x_i; \: \theta_f); \: \theta_p), \: y_i) \\ & - \lambda  \sum_{j=1}^{N_2} L_{d}(H_d(H_f (x_j; \: \theta_f); \: \theta_d), \: d_j) =  \\
    &= \sum_{i = 1}^{N_s} L_s \left(\hat{y}_i ,\: y_i \right) \: - \: \lambda \sum_{j = 1}^{N_2} L_d (\hat{d}_j,\: d_j )
    \end{split} \\ 
    (\hat{\theta}_f, \hat{\theta}_p) & = \argmin_{\theta_f, \theta_p} E(\theta_f, \theta_p, \theta_d), \\
    \hat{\theta}_d & = \argmax_{\theta_d} E(\theta_f, \theta_p, \theta_d),
\end{align}

where $L_s$ is the segmentation loss (BCE), $L_d$ is the domain classification loss (BCE). $\theta_f, \: \theta_p, \: \theta_d$ are the parameters of $H_f, \: H_p, \: H_d$, respectively, and $\hat{\theta}_f, \: \hat{\theta}_p, \: \hat{\theta}_d$ are the solutions we seek. The parameter $\lambda$ regulates the trade-off between the adversarial and segmentation objectives.

The goal of the discriminator is to classify a kernel that was used to reconstruct the image. To aggregate features before the discriminator, we use $1 \times 1$ convolutions and interpolation to equalize the number of channels and spatial size, then we concatenate the result. The discriminator consists of a sequence of fully-convolution layers followed by several fully-connected layers. We also use Leaky ReLU activations \citep{leaky_relu} and average pooling to avoid sparse gradients.

However, there is no consensus in the literature on how to connect the discriminator to a segmentation network \citep{zakazov2021anatomy}. Our experiments also show a high dependency of the model performance from the connection implementation. Therefore, we consider two strategies of connecting the discriminator: aggregating features from the earlier (encoder) and later (decoder) layers. We denote these approaches by \textit{DANN (Enc)} and  \textit{DANN (Dec)}, respectively. The \textit{DANN (Enc)} version is also presented in Fig. \ref{fig:method_schematic}. We describe the experimental details in Sec. \ref{ssec:exp:dann}.
% Instead of using final layers we propose to aggregate features from intermediate layers in a way similar to proposed in \cite{dou2018unsupervised} and \cite{kamnitsas2017unsupervised}.
% During our experiments we found out that discriminator model is unable to classify image domain based on outputs of the last layers. We suppose that at this stage all information about image style is lost.

\subsection{Cross-domain feature maps consistency}
\label{ssec:method:f-consistency}

Similarly to the adversarial approach, we propose to remove style-specific kernel information from the feature maps. However, we additionally exploit the paired nature of the unlabeled dataset $S_2$. Instead of the adversarial loss, we minimize the distance between feature maps of paired images. We use the same notations $H_f$, $H_p$, $\theta_f$, $\theta_p$, and $\hat{y} = H_p(H_f (x; \: \theta_f); \: \theta_p)$ as in Sec. \ref{ssec:method:dann}. Further, we denote the feature vector for every image $x$ as $f$, $f = H_f (x; \: \theta_f)$. For the paired image $\tilde{x}$, we use the similar notation $\tilde{f}$.

During the training, we minimize the sum of segmentation loss and distance between paired features ($f$ and $\tilde{f}$). Thus, the optimization problem is

\begin{align}
\label{eq:f-consistency_opt}
    \begin{split}
        E (\theta_f, \theta_p) &= \sum_{i = 1}^{N_s} L_s (H_p(H_f (x_i; \: \theta_f); \: \theta_p), \: y_i) \: \\
        & + \: \alpha \sum_{j = 1}^{N_2} L_c (H_f (x_j; \theta_{f}), \: H_f (\tilde{x}_j; \theta_f)) \\
        & = \sum_{i = 1}^{N_s} L_s \left(\hat{y}_i ,\: y_i \right) \: + \: \alpha \sum_{j = 1}^{N_2} L_c (f_j,\: \tilde{f}_j ),  
    \end{split} \\
    (\hat{\theta}_f, \hat{\theta}_p) & = \argmin_{\theta_f, \theta_p} E(\theta_f, \theta_p),
\end{align}
where $L_s$ is the segmentation loss (BCE) and $L_c$ is the \textit{consistency} loss. For the consistency loss, we use mean squared error (MSE) between paired feature maps. Parameter $\alpha$ regulates the trade-off between two objectives. We call this method \textit{F-Consistency} since it enforces the consistency between paired feature maps.
% The described method is depicted in Fig. \ref{fig:method_schematic}.

\begin{figure}[h]
    \centering
    \caption{Schematic representation of the proposed method, \textit{\textbf{F-Consistency}} (5), and its competitors, \textit{P-Consistency} (4) and \textit{DANN} (6). All methods build upon the same U-Net architecture, which we train to segment the COVID-19 binary mask (1) from the axial slices of chest CT images (2). These adaptation methods use unlabeled paired data (3) to improve the model performance on the target domain. We show the flow and different usage of the paired data in different methods with green. In the image, \textit{DANN} and \textit{F-Consistency} operate with the encoder layers but can be easily extended to the decoder versions.}
    \includegraphics[width=\textwidth]{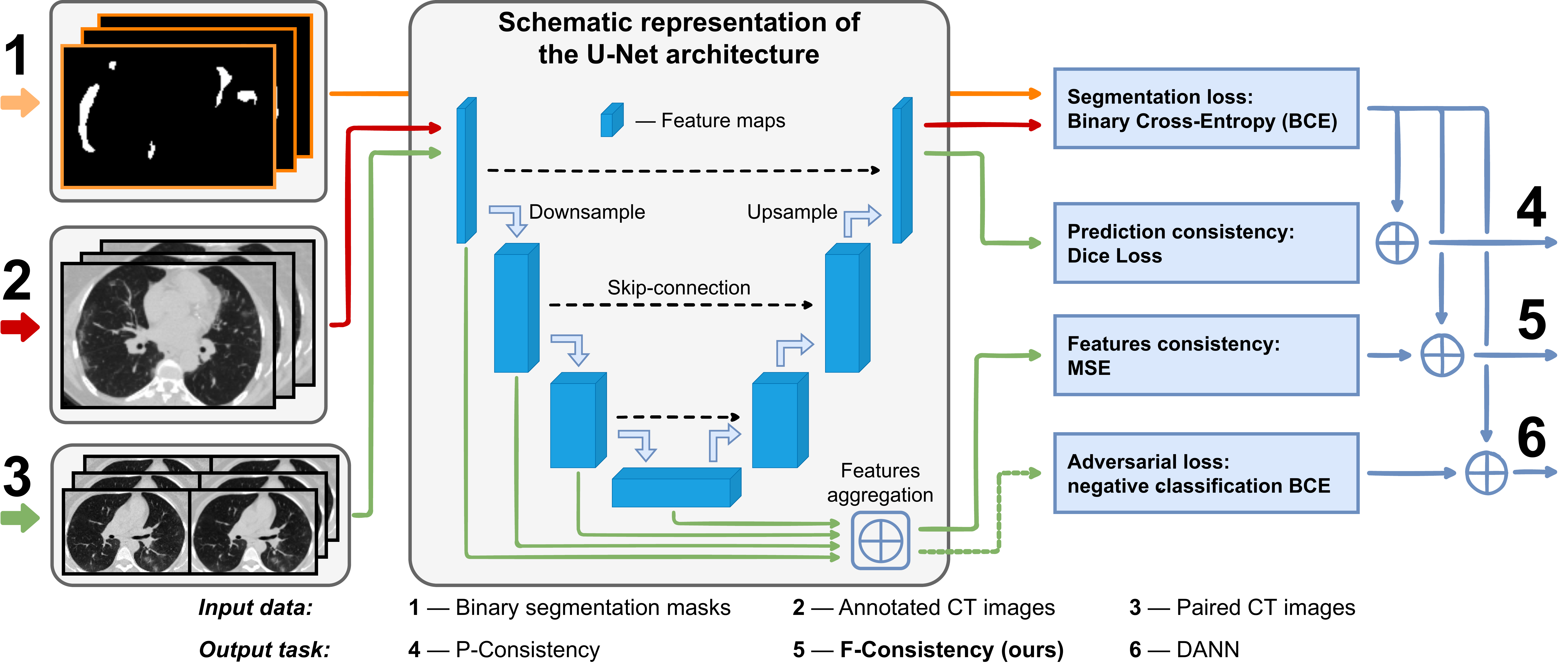}
    \label{fig:method_schematic}
\end{figure}

Along with DANN, we present our method schematically in Fig. \ref{fig:method_schematic} (5). Note that we do not need any additional model, e.g., discriminator $H_d$, in the case of F-Consistency. However, the same question of choosing the feature maps to aggregate arises. We consider the same two strategies as in Sec. \ref{ssec:method:dann}: aggregating encoder and decoder feature maps. We further call these implementations of our method \textit{F-Consistency (Enc)} and \textit{F-Consistency (Dec)}, respectively. All experimental details are given in Sec. \ref{ssec:exp:f-consistency}.

% In all experiments we use binary cross-entropy as supervised loss function $L_t$ and MSE-loss function as consistency loss $L_s$. During training stage we at every step use different mini-batches for segmentation and regularization tasks. For segmentation sub-task we sample a batch of images and their binary masks from the annotated dataset $S_s$. The second mini-batch consists of paired images sampled from the paired dataset $S_2$. We independently calculate loss function for both mini-batches. Then the losses are summed and model parameters are optimized during a single optimization step.

\subsection{Cross-domain predictions consistency}
\label{ssec:method:p-consistency}

A special case of F-Consistency (Dec) is enforcing the consistency of paired predictions, since the predictions are de facto the feature maps of the last network layer. This approach is proposed in \citep{orbes2019multi} also in the context of medical image segmentation. Further, we denote this method \textit{P-Consistency}. Visually, it could be compared with DANN and F-Consistency in Fig. \ref{fig:method_schematic} (4). The optimization problem is the same as in Eq. \ref{eq:f-consistency_opt}, except $L_c$ is Dice Loss \citep{vnet} and $f$ and $\tilde{f}$ are the last layer features, i.e., predictions. The experimental details are described in Sec. \ref{ssec:exp:p-consistency}.

% Such approach is much easier to implement since it penalizes final predictions and doesn't require additional step to extract intermediate features.  

% However, as it shown in Sec. \ref{ssec:res:public} this method has poor performance in setup where paired images do not contain lesions. We discuss this issues in Sec. \ref{ssec:res:public}.

\section{Data}
\label{sec:data}

In our experiments, we use a combination of different datasets with chest CT images. The data could be divided into two subsets according to the experimental purposes. The first collection of datasets consists of images with annotated COVID-19 lesions, i.e., with binary masks of ground-glass opacity and consolidation. It serves to train the COVID-19 segmentation algorithms. We detail every dataset of the segmentation collection in Sec. \ref{ssec:data:segm}. 

The second collection consists of chest CT images which are reconstructed with different kernels. We filter the data to contain pairs of images reconstructed with the \textit{smooth} and \textit{sharp} kernels. This data is further used to adapt the models in an unsupervised manner. We detail the second collection in Sec. \ref{ssec:data:paired}.

Most of these datasets are publicly available, yielding the reproducibility of our experiments. We summarize the description in Tab. \ref{tab:data_segm} and Tab. \ref{tab:data_paired}.

\subsection{Segmentation data}
\label{ssec:data:segm}

We use three publicly available datasets to train and test the segmentation algorithms: \textit{Mosmed-1110} \citep{morozov2020mosmeddata}, \textit{MIDRC} \citep{tsai2021rsna}, and \textit{Medseg-9}. We ensure that selected datasets contain original 3D chest CT imaging studies without third-party preprocessing artifacts. The images from the \textit{Mosmed-1110} and \textit{MIDRC} datasets are reconstructed using \textit{smooth} kernels, whereas \textit{Medseg-9} images have \textit{sharp} reconstruction kernels. That allows us firstly to identify and then address the domain adaptation setup. Therefore, we split the segmentation data into the source (\textit{COVID-train}) and target (\textit{COVID-test}) domains and describe them in Sec. \ref{sssec:data:segm:train} and Sec. \ref{sssec:data:segm:test}, respectively. Summary of the segmentation datasets is presented in Tab. \ref{tab:data_segm}.
% To the best of our knowledge, we cover the largest databases of COVID-19 cases that satisfy the reasonable conditions listed below.

\begin{table}[H]
    \centering
    \caption{Summary of the segmentation datasets.}
    \label{tab:data_segm}
    
    \resizebox{\textwidth}{!}{%
    \begin{tabular}{c c c c c c}
        \toprule
        Dataset & Source & \makecell{Effective \\ size} & Kernels & Annotations & Split \\
        \midrule
        \multirow{2}{*}[-1.5em]{\textit{COVID-train}} & \makecell{\textit{Mosmed-1110} \\ \citep{morozov2020mosmeddata}} & 50 & \makecell{unknown \\ \textit{smooth}} & COVID-19 mask & \multirow{2}{*}[-1.5em]{\makecell{5-fold \\ cross-val}} \\
        \cmidrule(lr){2-5}
         & \makecell{\textit{MIDRC} \\ \citep{tsai2021rsna}} & 112 & \makecell{B/L/BONE/ \\ STANDARD \\ (\textit{smooth})} & COVID-19 mask & \\
        \midrule
        \textit{COVID-test} & \textit{Medseg-9} & 9 & \makecell{unknown \\ \textit{sharp}} & \makecell{COVID-19 mask, \\ lungs mask} & \makecell{hold-out \\ test} \\
        \bottomrule
        
    \end{tabular}}
\end{table}

\subsubsection{COVID-train}
\label{sssec:data:segm:train}

\paragraph{Mosmed-1110} This dataset consists of $1110$ chest CT scans collected in Moscow clinics during the first months of $2020$ \citep{morozov2020mosmeddata}. Scanning was performed on \textit{Canon (Toshiba) Aquilion 64} units using standard scanner's protocol: inter-slice distance of $0.8$ mm and \textit{smooth} reconstruction kernels in particular. However, the public version of \textit{Mosmed-1110} contains every $10th$ slice of the original series, which makes the resulting slice distance equal to $8.0$ mm.

Additionally, $50$ series have annotated binary masks depicting COVID-19 lesions (ground-glass opacity and consolidation). Further, we use only these $50$ images in our experiments. Also, as one of the preprocessing steps, we crop images to lung masks. However, lungs are not annotated in the dataset. We obtain the lung masks using a standalone algorithm; see details in Sec. \ref{ssec:exp:covid_segm}.

\paragraph{MIDRC} \textit{MIDRC-RICORD-1a} is the publicly available dataset that contains $120$ chest CT studies \citep{tsai2021rsna}. The total number of volumetric series is $154$. According to the DICOM entries, most images have \textit{smooth} reconstruction kernels. The dataset contains at least $12$ paired images (without considering the studies that contain more than two series). However, we do not use these pairs to enforce consistency since both images have \textit{smooth} kernels. Also, the original dataset does not contain annotated lung masks. Therefore, similarly to \textit{Mosmed-1110}, we predict the lung masks with a separate model (Sec. \ref{ssec:exp:covid_segm}). % All studies were performed using \textit{Phillips} scanner.

Finally, we use only the images that have non-empty annotations. The images that have empty binary masks of COVID-19 are discarded both from the \textit{Mosmed-1110} and \textit{MIDRC} datasets. The resulting training dataset consists of $112$ volumetric images with \textit{smooth} kernels.

\subsubsection{COVID-test}
\label{sssec:data:segm:test}

\paragraph{Medseg-9}
MedSeg website\footnote{\url{https://medicalsegmentation.com/covid19/}} shares a publicly available dataset with $9$ annotated chest CT images from \textit{here}\footnote{\url{https://radiopaedia.org/articles/covid-19-3}}. Although there is no information about reconstruction kernels, we perceptually identify these images to have \textit{sharp} kernels. The latter assumption also finds an experimental confirmation in the significantly lower scores when the model is trained on the images with \textit{smooth} kernels. Contrary to the \textit{COVID-train} dataset, \textit{Medseg-9} contains annotated lung masks. However, we find the masks predicted by our algorithm more precise and use the predicted ones.% All images visually correspond to an edge preserving convolutional kernels.

Also, we ignore the other dataset with $20$ annotated CT images from MedSeg website \citep{external20}. The preprocessing of these images is unknown, inconsistent, and diverges from our default preprocessing pipeline. Therefore, our testing dataset consists of $9$ images with \textit{sharp} kernels.

\subsection{Paired images data}
\label{ssec:data:paired}

To train and evaluate the consistency of the segmentation algorithms, we use two sources of paired data. The first source is a publicly available dataset \textit{Cancer-500} \citep{cancer500}. However, the \textit{Cancer-500} dataset does not contain COVID-19 cases. Therefore, to properly evaluate the consistency of COVID-19 segmentation algorithms, we use the second source of private data that contains COVID-19 cases. 

From \textit{Cancer-500}, we build the \textit{Paired-public} dataset (Sec. \ref{sssec:data:paired:public}) and use it only to train the segmentation algorithms in an unsupervised manner. Then, we build the \textit{Paired-private} dataset from our private data (Sec. \ref{sssec:data:paired:private}). Besides training, we use this dataset to evaluate the consistency scores since it contains the segmentation target (COVID-19 lesions). The summary of the paired datasets is presented in Tab. \ref{tab:data_paired}.

Both datasets do not contain any COVID-19 or lungs annotations. Note that we do not need COVID-19 annotations since we use these datasets in an unsupervised training. However, we need lung masks to preprocess images. Thus, we use the same lungs segmentation model (Sec. \ref{ssec:method:covid_segm}) as for the other datasets.

\begin{table}[H]
    \centering
    \caption{Summary of the datasets with paired images.}
    \label{tab:data_paired}
    
    \resizebox{\textwidth}{!}{%
    \begin{tabular}{c c c c}
    \toprule
    Dataset & Kernel pair (\textit{smooth}/\textit{sharp}) & Training & Testing pairs \\
    \midrule
    \multirow{2}{*}{\makecell{Paired-public \\ \citep{cancer500}}} & FC07/FC55 & 22 & 0 \\
     & FC07/FC51 & 98 & 0 \\
    \midrule
    \multirow{4}{*}{Paired-private} & FC07/FC55 & 60 & 20 \\
     & FC07/FC51 & 30 & 11 \\
     & SOFT/LUNG & 30 & 10 \\
     & STANDARD/LUNG & 30 & 10 \\
    \bottomrule
\end{tabular}}

\end{table}

\subsubsection{Paired-public}
\label{sssec:data:paired:public}

We build the \textit{Paired-public} dataset using a publicly available dataset, \textit{Cancer-500} \citep{cancer500}. The data was collected from $536$ randomly selected patients of Moscow clinics in 2018. All original images were obtained using a \textit{Toshiba} scanner and reconstructed with FC07, FC51, or FC55 kernels. Here, FC07 is a \textit{smooth} reconstruction kernel, whereas FC51 and FC55 are \textit{sharp} kernels. From $536$ studies, we extracted $120$ pairs, comparing the shape and acquisition time of the corresponding DICOM series and filtering contrast-enhanced cases. As a result, the \textit{Paired-public} dataset consists of $98$ FC07/FC51 and $22$ FC07/FC55 pairs (Tab. \ref{tab:data_paired}). We use this dataset to train the domain adaptation algorithms on paired images. % 970 series. Details are provided in Tab.

However, the \textit{Paired-public} dataset does not contain COVID-19 cases (it was collected before the pandemic). The latter observation limits using this dataset to evaluate the consistency. Otherwise, we evaluate the quality of COVID-19 segmentation algorithms using images with no COVID-19 lesions. Thus, we either evaluate the consistency of noisy or false positive predictions. For the same reason, one should also be careful using this data to enforce the consistency in the last network layers, e.g., in \textit{P-Consistency} (Sec. \ref{ssec:method:p-consistency}). The data without COVID-19 lesions can force the network to output trivial predictions.

Therefore, we introduce a private dataset for the extended consistency evaluation and robust training of some of the domain adaptation algorithms.

\subsubsection{Paired-private}
\label{sssec:data:paired:private}

From the private collection of the chest CT images, we filter out $181$ pairs to create the \textit{Paired-private} dataset. These images were initially collected from Moscow outpatient clinics during the year 2020. Scanning was performed on the \textit{Toshiba} and \textit{GE medical systems} units using diverse settings. We select the four most frequent kernel pairs with a total of six unique reconstruction kernels: FC07, FC51, FC55, LUNG, SOFT, and STANDARD. We detail the distribution of kernel pairs in Tab. \ref{tab:data_paired}.

Due to the purpose of collecting the data, these images contain COVID-19 lesions. Therefore, we use the \textit{Paired-private} dataset both for training and evaluation. The wider variety of kernels also allows us to test the generalization of algorithms to unseen kernels; see further experimental details in Sec. \ref{sec:exp}.

\section{Experiments}
\label{sec:exp}

The main focus of experiments below is to compare our method to the other unsupervised domain adaptation techniques. To achieve an objective comparison, a fair and unified experimental environment should be created. Therefore, we firstly describe the common preliminary steps that build up every method. This description includes preprocessing, lungs segmentation, and COVID-19 segmentation steps in Sec. \ref{ssec:exp:covid_segm}.
% the similar network architectures, training procedures, etc.

Further, we detail every of the domain adaptation methods: Filtered Backprojection Augmentation (FBPAug) in Sec. \ref{ssec:exp:fbpaug}, Deep Adversarial Neural Network in Sec. \ref{ssec:exp:dann}, \textit{cross-domain feature-maps consistency} which is our proposed method in Sec. \ref{ssec:exp:f-consistency}, and cross-domain predictions consistency in Sec. \ref{ssec:exp:p-consistency}.

In all latter experiments, we use the same data split and evaluation metrics. Firstly, we split the \textit{COVID-train} dataset (source domain with smooth kernels) into $5$ folds. Then, we perform a standard cross-validation procedure, training on the data from four folds and calculating the score on the remaining fold. Here, we calculate the Dice Score between the predicted and ground truth COVID-19 masks for every 3D image and average these scores for the whole fold. Also, for every validation, we calculate the average Dice Score on the \textit{COVID-test} dataset, which is the target domain with sharp kernels. Finally, we report the mean and standard deviation of these five scores on cross-validation and target domain data.

Besides the Dice Score on the source and target data, we also report the Dice Score between predictions on the paired images. To do so, we split the \textit{Paired-private} datasets' pairs into training and testing folds stratified by the type of kernel pairs. The size of the test fold is approximately $30\%$ of the dataset size. Then, we supplement the source domain training data (four current folds of the cross-validation) with the fixed training part of the paired data. Average Dice Score is calculated on the test part in a similar fashion.

% \todo{We summarize the splits in Tab. N.}

\subsection{COVID-19 segmentation}
\label{ssec:exp:covid_segm}

\paragraph{Preprocessing}
We use the same preprocessing steps for all experiments. Firstly, we rescale all CT images to have $1.75 \times 1.75$ mm axial resolution. Then, the intensity values are clipped to the minimum of $-1000$ Hounsfield units (HU) and maximum of $300$ HU. The resulting intensities are min-max-scaled to the $[0; 1]$ range.

\paragraph{Lungs segmentation}
Further, we crop CT images to a bounding box of the lungs mask. We obtain the latter mask by training a standalone CNN segmentation model. The training procedure and architecture are reproduced from \citep{goncharov2021ct}. The training of the lung segmentation model involves two external chest CT datasets: LUNA16 \citep{luna16,armato2011lung} and NSCLC-Radiomics \citep{nsclc1,nsclc2}. These datasets have an empty intersection with the other datasets used to train the COVID-19 segmentation models; thus, there is no leak of the test data.

\paragraph{COVID-19 segmentation}
In all COVID-19 segmentation experiments, we use the same 2D U-Net architecture described in Sec. \ref{ssec:method:covid_segm}. We train all models for $25$k iterations using Adam \citep{adam} optimizer with the default parameters and an initial learning rate of $10^{-4}$. Every $6$k batches learning rate is multiplied by $0.2$. Each iteration contains $32$ randomly sampled 2D axial slices. Training of the segmentation model takes approximately $12$ hours on nVidia GTX $1080$ ($8$ GB).
% half-precision

We further call the model trained only on a source data and without any pipeline modifications a \textbf{baseline}. For the baseline, we also calculate test scores on the \textit{COVID-test} dataset and consistency scores on the \textit{Paired-private} dataset. We refer to it as a starting point for all other methods.

\subsection{Filtered Backprojection Augmentation}
\label{ssec:exp:fbpaug}

The first method that we consider as a solution to our domain shift problem is FBPAug \citep{saparov2021zero}. One can find the relevant method description and motivation to use it in Sec. \ref{ssec:method:fbpaug}. In our experiments, we use the original implementation of FBPAug from \citep{saparov2021zero} and also preserve the augmentation parameters. However, we sample parameters from the interval that corresponds to \text{sharp} reconstruction kernels ($a$ from $[10.0, 40.0]$, $b$ from $[1.0, 4.0]$), since our goal is to adapt model to sharp kernels. We also reduce the probability of augmenting an image from $0.5$ to $0.1$. The latter change does not affect performance (tested on a single validation fold) and reduces the experiment time (saving about $90$ hours per experiment).

Note that the experimental setup remains the same as in baseline (Sec. \ref{ssec:exp:covid_segm}). FBPAug is the only adaptation method that does not use paired data.

\subsection{Deep Adversarial Neural Network}
\label{ssec:exp:dann}

The next step is to use unlabeled paired data to build a robust to domain shift algorithm. Here, we adopt a DANN approach \citep{uda} to the COVID-19 segmentation task. We detail this approach in Sec. \ref{ssec:method:dann}. 

In our experiments, we use the scheduling of parameter $\lambda$ as in \citep{uda}. The baseline training procedure is extended to sample from the unlabeled data. At every iteration, we additionally sample $16$ pairs of axial slices (the batch size is $32$) from one of the \textit{Paired-private} or \textit{Paired-public} datasets (depending on a data setup). Then, we make the second forward pass to the discriminator and sum the segmentation and adversarial losses. The rest of the pipeline remains the same as in the baseline.

For this method, we select two parameters that can drastically change its behaviour in terms of consistency and segmentation quality. Firstly, we evaluate different $\lambda$ values, where $\lambda$ determines how strongly adversarial loss contributes to the total loss (see Sec. \ref{ssec:method:dann}). With the close to zero $\lambda$, we expect DANN to behave similar to baseline. With the larger $\lambda$, we expect our segmentation model starting to fool the discriminator, making features of the different kernels indistinguishable for the discriminator. However, this consequence does not guarantee the increase of consistency or segmentation quality. Therefore, we manually search for the $\lambda \in \{10^{-5}, 10^{-4}, 10^{-3}, 10^{-2}, 10^{-1}, 1\}$ and choose the best. Secondly, as discussed in Sec. \ref{ssec:method:dann}, we compare two approaches to connecting the discriminator to the segmentation model: (i) to the encoder and (ii) to the decoder part.

Finally, we evaluate how well DANN generalizes under the presence of different kernel pairs. To do so, we exclude SOFT/LUNG and STANDARD/LUNG kernel pairs from training. We compare the results of this experiment with the model that is trained on all available kernel pairs from \textit{Paired-private}. We also test the sensitivity of the DANN approach to the presence of COVID-19 lesions in the unlabeled data. In this case, we train DANN on the \textit{Paired-public} dataset that does not contain COVID-19 targets.

\subsection{Cross-domain feature maps consistency}
\label{ssec:exp:f-consistency}

Our proposed F-Consistency also uses unlabeled paired data. Therefore, the training procedure is the same as for DANN (Sec. \ref{ssec:exp:dann}), except we do not use any scheduling for parameter $\alpha$. Our method is detailed in Sec. \ref{ssec:method:f-consistency}.

Similarly to DANN's experimental setup, we select two parameters to evaluate: different $\alpha$ values and the features that contribute to the consistency regularization. Here, $\alpha$ controls the trade-off between the consistency and segmentation quality. However, contrary to the discriminator's $\lambda$ in Sec. \ref{ssec:exp:dann}, the large $\alpha$ values for consistency regularization ensures the features alignment. We show this trade-off for ten $\alpha$ values in a log-space from $3^{-10}$ to $1$. Then, we compare two approaches to enforcing the features consistency: regularizing encoder's and decoder's features.

Finally, we evaluate the generalization of F-Consistency to different kernel pairs similarly by excluding SOFT/LUNG and STANDARD/LUNG kernel pairs from training. Similarly to DANN, we train our method on the \textit{Paired-public} dataset that does not contain COVID-19 lesions and show its generalization to kernel styles, regardless of semantic content.

\subsection{Cross-domain predictions consistency}
\label{ssec:exp:p-consistency}

One special case of F-Consistency is enforcing the paired predictions consistency, which is independently evaluated in \citep{orbes2019multi}. We call this case a P-Consistency and detail it in Sec. \ref{ssec:method:p-consistency}.

We follow the experimental setup as in F-Consistency (Sec. \ref{ssec:exp:f-consistency}). We show the trade-off between the target and consistency scores for the same values of $\alpha$. However, there are no experiments on regularizing specific features (in encoder or decoder), since predictions consistency can be regularized only at the output network's layer.

For the P-Consistency, we draw one's attention to the experiment on the \textit{Paired-public} dataset. Since this dataset does not contain COVID-19 cases, the enforced predictions consistency on empty-target images can result in trivial predictions. Thus, we expect P-Consistency to be less generalizable to the target domain in terms of target Dice Score. However, for the datasets with COVID-19 lesions we show the generalization for the unseen kernels as for the other methods.
% By using paired images without COVID lesions we are trying to investigate the ability of models to extract style information only, regardless of semantic content.

\section{Results}
\label{sec:results}

We structure our experimental results as follows. We firstly compare the final versions of all methods in Sec. \ref{ssec:res:final} so we directly support our main message. Secondly, we compare the generalization of all methods trained on less data in Sec. \ref{ssec:res:generalization}. Then, we visualize the trade-off between the consistency and COVID-19 segmentation quality in Sec. \ref{ssec:res:trade-off}.

The experimental methodology follows Sec. \ref{sec:exp}. We compare some of the key results statistically using one-sided Wilcoxon signed-rank test. We report p-values in two testing setups: $p_1$, comparing five mean values after cross-validation, and $p_2$, comparing Dice Score on every image as an independent sample.

\subsection{Methods comparison}
\label{ssec:res:final}

\begin{table}[H]
    \centering
    \caption{Comparison of the best versions of all considered methods from Sec \ref{sec:method}. The adaptation methods are trained using all training kernel pairs of the \textit{Paired-private} dataset. \textit{F-}/\textit{P-Cons} stand for F-/P-Consistency, where F-Consistency is our proposed method. \textit{Enc} and \textit{Dec} correspond to the implementation of methods operating with the encoder and decoder network layers, respectively. All results are Dice Scores in the format \textit{mean} $\pm$ \textit{std} calculated from $5$-fold cross-validation. We highlight the best scores in every column in \textbf{bold}.}
    %\vspace{.2cm}
    \resizebox{\textwidth}{!}{%
    \begin{tabular}{l c c c c c c c}
        \toprule
        
        & \multirow{2}{*}[-0.2em]{\textit{COVID-train}} & \multirow{2}{*}[-0.2em]{\textit{COVID-test}} & \multicolumn{5}{c}{\textit{Paired-private} consistency} \\
        \cmidrule(lr){4-8}
        & & & FC07/55 & FC07/51 & SOFT/LUNG & STAND/LUNG  & Mean \\

        \midrule
        Baseline & $.60 \pm .04$ & $.56 \pm .03$ & $.52 \pm .06$ & $.39 \pm .07$ & $.58 \pm .08$ & $.28 \pm .05$ & $.46 \pm .05$\\
        
        \cmidrule{1-8}
        
        FBPAug & $.59 \pm .04$ & $.62 \pm .03$ & $.80 \pm .02$ & $.71 \pm .03$ & $.85 \pm .01$ & $.65 \pm .03$ & $.76 \pm .02$ \\

        \cmidrule{1-8}
        
        DANN (Dec) & $.57 \pm .04$ & $.61 \pm .04$ &  $.61 \pm .02$ & $.49 \pm .04$ & $.58 \pm .03$ & $.31 \pm .05$ & $.52 \pm .01$ \\
        
        \cmidrule{1-8}
         
        DANN (Enc) & $.58 \pm .05$ & $\mathbf{.64 \pm .02}$ & $.84 \pm .02$ & $.70 \pm .02$ & $\mathbf{.86 \pm .03}$ & $.66 \pm .06$ & $.78 \pm .02$ \\
        
        \cmidrule{1-8}
        
        P-Cons & $.59 \pm .04$ & $.61 \pm .01$ & $.65 \pm .05$ & $.60 \pm .02$ & $.77 \pm .01$ & $.47 \pm .04$ & $.63 \pm .03$\\
        
        \cmidrule{1-8}

        F-Cons (Dec) & $\mathbf{.60 \pm .03}$ & $.58 \pm .02$ & $.62 \pm .05$ & $.54 \pm .03$ & $.75 \pm .01$ & $.40 \pm .06$ & $.58 \pm .02$ \\
    
        \cmidrule{1-8}
        
        F-Cons (Enc) &  $.57 \pm .03$ & $\mathbf{.64 \pm .03}$ & $\mathbf{.88 \pm .01}$ & $\mathbf{.72 \pm .04}$ & $.83 \pm .02$ & $\mathbf{.70 \pm .05}$ & $\mathbf{.80 \pm .01}$ \\
         
    \bottomrule
    \end{tabular}}
    \label{tab:res_final}
\end{table}

% we analyze the quality of proposed method and compare it to other methods describe in Sec. \ref{sec:method} in setup where all types of kernels pairs are available during train stage.

To begin with, we show the existence of the domain shift problem within the COVID-19 segmentation task. The Dice Score of the baseline model on the \textit{COVID-test} dataset is lower than the cross-validation score on the \textit{COVID-train} dataset, $0.56$ against $0.60$. Also, this score is significantly lower than $0.64$ achieved by our adaptation method ($p_1 < 0.05$, $p_2 < 10^{-4}$). One could find the results in Tab. \ref{tab:res_final} comparing row \textit{Baseline} to the others. For all adaptation methods, we observe an increase in the consistency score and segmentation quality on the target domain. Moreover, all methods maintain their quality on the source domain comparing to \textit{Baseline}.

Further, we compare \textit{FBPAug} to the best adaptation methods since it is a straightforward solution to the domain shift problem caused by the difference in the reconstruction kernels (Sec. \ref{ssec:method:fbpaug}). Although \textit{FBPAug} achieves comparable results on the target domain, our method outperforms it in terms of the average consistency score, $0.80$ Dice Score against $0.76$ ($p_1 < 0.05$, $p_2 < 10^{-5}$). The results are also in Tab. \ref{tab:res_final}, row \textit{FBPAug}.

We highlight two possible reasons for FBPAug to perform worse than its competitors. Firstly, the augmented data is synthetic since the method emulates but does not reproduce the CT reconstruction process. Secondly, the range of augmentation is limited compared to the diversity of the real reconstructed images. Therefore, the lower quality and diversity of the augmented data might affect the performance of FBPAug, resulting in lower consistency scores.
% Furthermore, in Sec. \ref{ssec:res:public}, we show our method to degrade and perform similar to FBPAug when trained with the limited paired data.

Finally, we compare the models that depend on paired regularization data: DANN, F-Consistency (our proposed method), and P-Consistency; the last five rows in Tab. \ref{tab:res_final}.

\textit{Implementing methods that operate with the encoder layers outperforms those with the decoder layers.} For the adversarial approach, \textit{DANN (Dec)}'s consistency score $0.52$ is significantly lower than $0.77$ of \textit{DANN (Enc)} ($p_1 < 0.05$, $p_2 < 10^{-10}$). The same is true comparing \textit{F-Cons (Dec)} and \textit{F-Cons (Enc)}: consistency score $0.58$ is significantly lower than $0.80$ ($p_1 < 0.05$, $p_2 < 10^{-10}$). We also note that P-Consistency operates with the last \textit{decoder} layer; thus, we compare it with \textit{F-Cons (Enc)} and show it resulting in the significantly lower consistency score, $0.63$ against $0.80$ ($p_1 < 0.05$, $p_2 < 10^{-10}$). Thus, our experimental evidences align with the message of \citep{zakazov2021anatomy} that the earlier (encoder) layers contain more domain-specific information than the later (decoder, output) ones.

\textit{Besides F-Consistency outperforms P-Consistency, we also show our method to outperform DANN (encoder versions).} Although both methods score similar in terms of target Dice Score, F-Consistency has an advantage over DANN in the consistency score: $0.80$ against $0.78$ ($p_1 < 0.05$, $p_2 < 10^{-2}$). Our intuition here is F-Consistency explicitly enforces features alignment and DANN enforces features to be indistinguishable for the discriminator. The latter differently impacts on the consistency score, and F-Consistency performs better.

\begin{figure}[h]
    \caption{Examples of axial CT slices from the \textit{COVID-test} dataset with the corresponding predictions and ground truth annotations. Three columns, denoted A, B, and C, contains three unique slices. The top row contains contours of the ground truth and baseline prediction. The bottom row contains contours of the adaptation methods' predictions. DANN and F-Consistency correspond to \textit{DANN (Enc)} and \textit{F-Cons (Enc)} from Tab. \ref{tab:res_final}, respectively.}
    \label{fig:test_preds}
    \includegraphics[width=\textwidth]{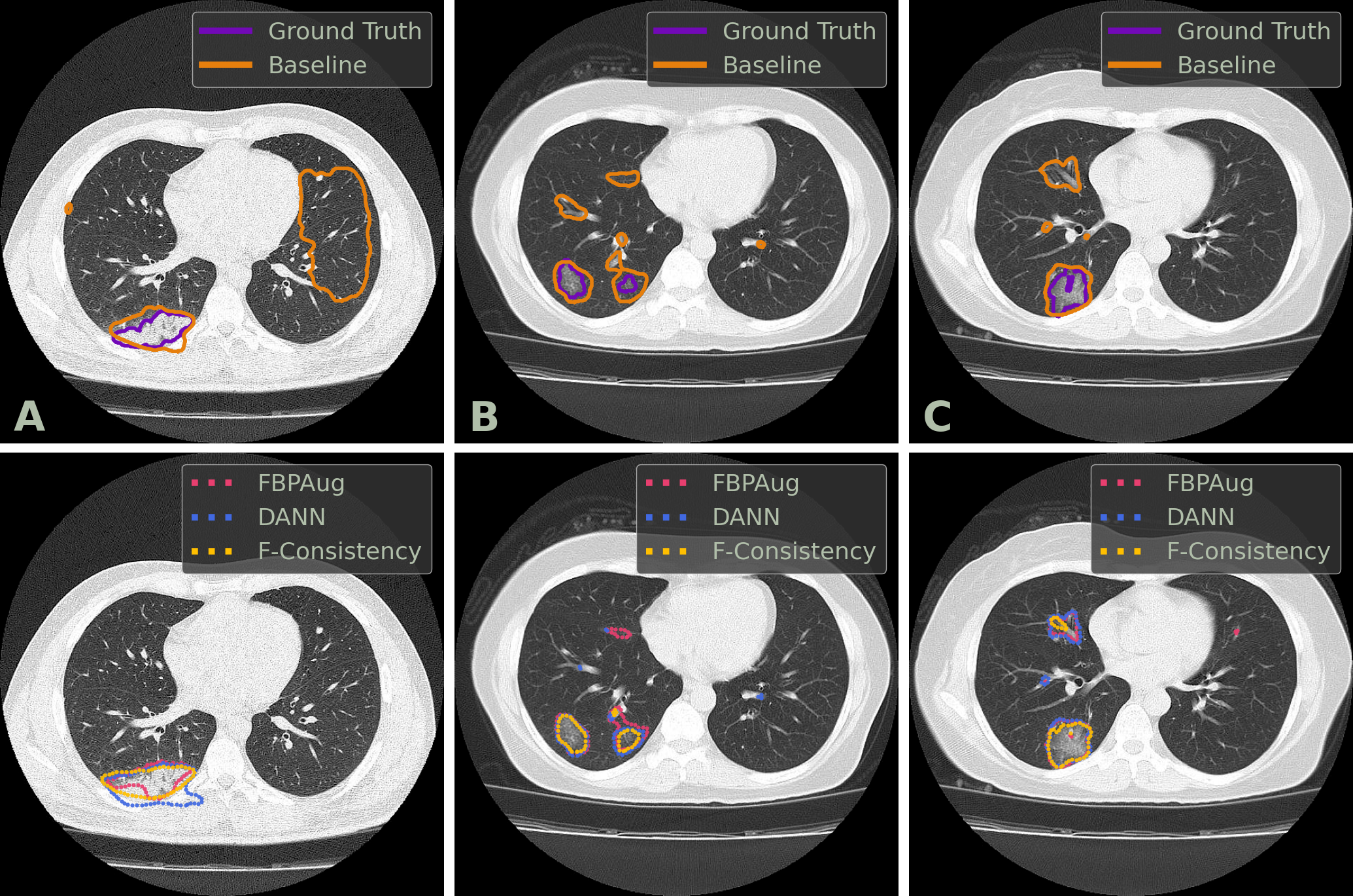}
    \centering
\end{figure}

\begin{figure}[h]
    \caption{Examples of CT slices from the \textit{Private-paired} dataset with the corresponding predictions on the paired images. Four doublets, denoted A, B, C, and D, contain corresponding slices from the \textit{smooth} and \textit{sharp} images. The doublets B and D are coronal and sagittal slices, respectively; we also pad them to visually align with the axial slices. Every slice contains predictions of four methods named in the legend. \textit{DANN} and \textit{F-Consistency} correspond to \textit{DANN (Enc)} and \textit{F-Cons (Enc)} from Tab. \ref{tab:res_final}, respectively.}
    \label{fig:consistency_preds}
    \includegraphics[width=\textwidth]{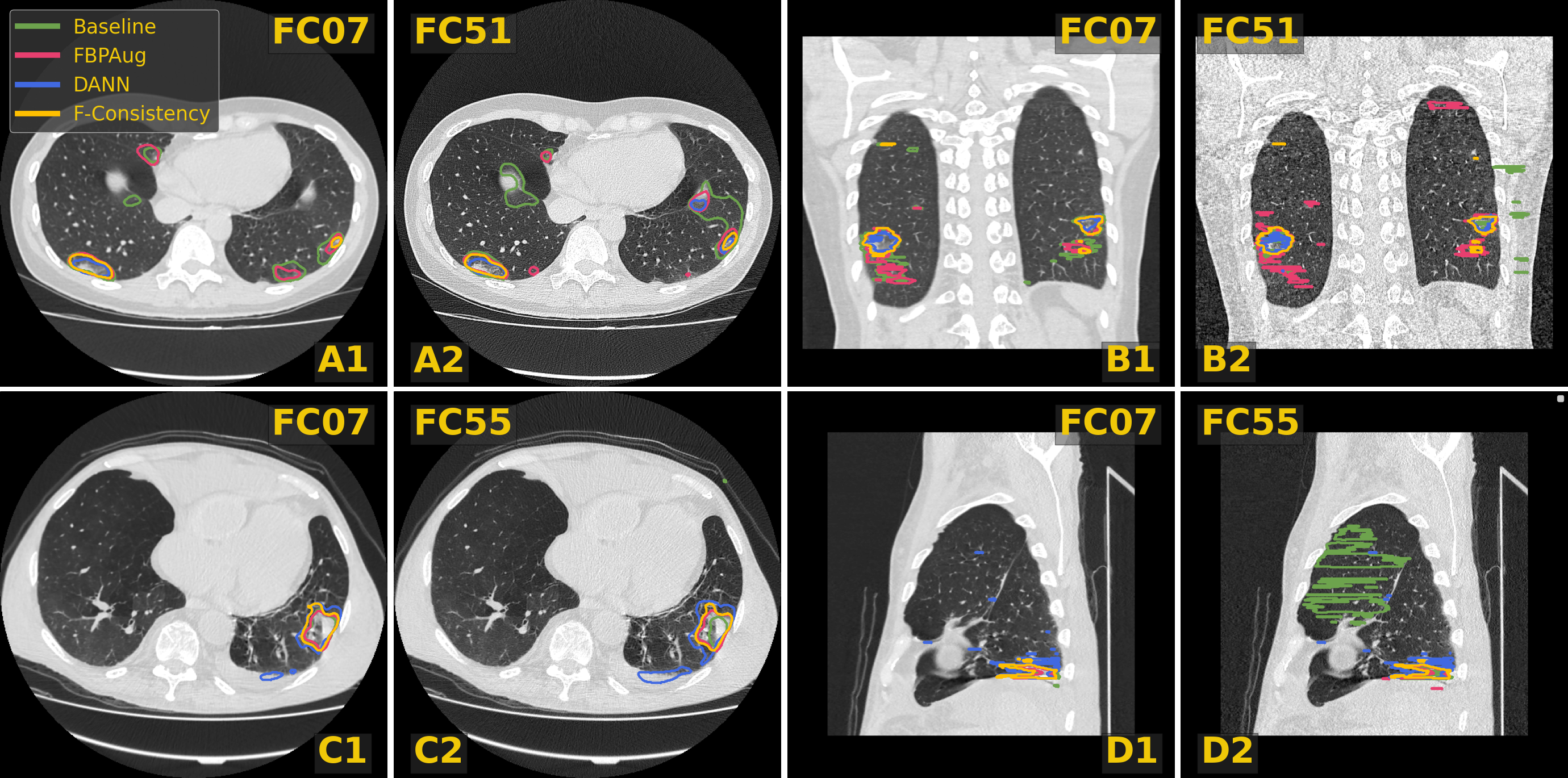}
    \centering
\end{figure}

We conclude the comparison of the methods comparison with the qualitative analysis. In Fig. \ref{fig:test_preds}, one could find examples of the \textit{Baseline}, \textit{FBPAug}, \textit{DANN (Enc)}, and \textit{F-Cons (Enc)} predictions on the \textit{COVID-test} dataset and compare them with the ground truth. Although all adaptation methods perform similar to the ground truth with minor inaccuracies, \textit{Baseline} outputs the massive false positive predictions on the unseen domain. Additionally to the quantitative analysis above, the latter observation highlights the relevance of the domain adaptation problem in the COVID-19 segmentation task.

In Fig. \ref{fig:consistency_preds}, we visualize predictions of the same four methods on the paired images from the \textit{Paired-private} dataset. For the \textit{Baseline}, we observe an extreme inconsistency (Fig. \ref{fig:consistency_preds}, A) and massive false positive predictions in healthy lung tissues (Fig. \ref{fig:consistency_preds}, D) and even outside lungs (Fig. \ref{fig:consistency_preds}, B). For the adaptation methods, their predictions are visually more consistent inside every pair, which aligns with the consistency scores in Tab. \ref{tab:res_final}.  Despite the high consistency scores, \textit{FBPAug} and \textit{DANN} output perceptually more aggressive predictions. \textit{FBPAug} predicts motion artifacts near the body regions (Fig. \ref{fig:consistency_preds}, A) and triggers similarly as the baseline on, most likely, healthy lung tissues (Fig. \ref{fig:consistency_preds}, B). \textit{DANN} is more conservative but triggers on the consolidation-like tissues (Fig. \ref{fig:consistency_preds}, C and D). However, without the ground truth annotations on the paired data, we refer to this analysis as a discussion. 
% The important thing here is nonzero preds.

Below, we investigate the generalization of the models trained with the less data and trade-off between consistency and segmentation quality.

\subsection{Generalization with the less data}
\label{ssec:res:generalization}

Firstly, we show how DANN, P-Consistency, and F-Consistency generalize to the unseen reconstruction kernels. We remove SOFT/LUNG and STANDARD/LUNG pairs of the \textit{Paired-private} dataset from training, so we train the models using FC07/FC51 and FC07/FC55 pairs. The results on the removed kernel pairs are shown in Tab. \ref{tab:res_kernels_gen}.

The methods preserve their segmentation quality on the \textit{COVID-train} and \textit{COVID-test} datasets despite we train them with limited data. Moreover, all three methods score considerably higher than \textit{Baseline} in consistency scores for unseen kernel pairs (SOFT/LUNG and STANDARD/LUNG). The latter means that the adaptations methods manage to align stylistic-related features even from the limited number of training examples. However, we highlight a decrease of the average consistency scores comparing to the versions trained on full \textit{Paired-private}. At this point, FBPAug (Tab. \ref{tab:res_final}) outperforms the adaptation methods. The latter indicates that the range of synthetically augmented data overlaps the range of reduced \textit{Paired-private}.

\begin{table}[h]
    \centering
    \caption{Comparison of \textit{DANN (Enc)}, \textit{P-Consistency}, and \textit{F-Consistency (Enc)} generalizing to previously unseen SOFT, STANDARD, and LUNG kernels. The numbers in the brackets next to the methods correspond to the number of kernel pairs in the \textit{Paired-private} dataset they are trained with, e.g., \textit{DANN (4)} matches with the \textit{DANN (Enc)} in Tab. \ref{tab:res_final}. All results are Dice Scores in the format \textit{mean} $\pm$ \textit{std} calculated from 5-fold cross-validation.}
    \resizebox{\textwidth}{!}{%
    \begin{tabular}{l c c c c c c c}
        \toprule
        
         & \multirow{2}{*}[-0.2em]{\shortstack{\textit{COVID-train}}} & \multirow{2}{*}[-0.2em]{\textit{COVID-test}}& \multicolumn{5}{c}{\textit{Paired-private} consistency} \\
        \cmidrule(lr){4-8}
         &  & & FC07/55 & FC07/51 & SOFT/LUNG & STAND/LUNG  & Mean\\
        
        \midrule
       Baseline & $.60 \pm .04$ & $.56 \pm .03$ & $.52 \pm .06$ & $.39 \pm .07$ & $.58 \pm .08$ & $.28 \pm .05$ & $.46 \pm .05$\\
        
        \cmidrule{1-8}
        
        % DANN (Dec) & $.58 \pm .04$ & $.62 \pm .05$ & $.639 \pm .015$ & $.45 \pm .04$ & $.58 \pm .05$ & $.30 \pm .03$ & $.52 \pm .02$\\
        
        % \cmidrule{1-8}
        
        DANN (4) & $.58 \pm .05$ & $.64 \pm .02$ & $.84 \pm .02$ & $.70 \pm .02$ & $.86 \pm .03$ & $.66 \pm .06$ & $.78 \pm .02$ \\
        
        % \cmidrule{2-8}
         
        DANN (2) & $.59 \pm .05$ & $.64 \pm .02$ & $.81 \pm .03$ & $.70 \pm .03$ & $.74 \pm .02$ & $.58 \pm .07$ & $.73 \pm .02$ \\
        
        \cmidrule{1-8}
        
        P-Cons (4) & $.59 \pm .04$ & $.61 \pm .01$ & $.65 \pm .05$ & $.60 \pm .02$ & $.77 \pm .01$ & $.47 \pm .04$ & $.63 \pm .03$\\
        
        % \cmidrule{1-8}
          
        P-Cons (2) &  $.59 \pm .04$ &  $.59 \pm .03$ & $.62 \pm .03$ & $.56 \pm .02$ & $.72 \pm .01$ & $.40 \pm .04$ & $.59 \pm .02$ \\
        
        \cmidrule{1-8}
        
        % F-cons (Dec) & $.59 \pm .04$ & $.57 \pm .02$ & $.59 \pm .06$ & $.51 \pm .03$ & $.69 \pm .03$ & $.34 \pm .05$ & $.54 \pm .02$ \\  
         
        % \cmidrule{1-8}
        
        F-Cons (4) &  $.57 \pm .03$ & $.64 \pm .03$ & $.88 \pm .01$ & $.72 \pm .04$ & $.83 \pm .02$ & $.70 \pm .05$ & $.80 \pm .01$ \\
        
        % \cmidrule{1-8}
          
        F-Cons (2) & $.58 \pm .04$ & $.64 \pm .01$ & $.83 \pm .02$ & $.64 \pm .03$ & $.75 \pm .02$ & $.59 \pm .02$ & $.73 \pm .01$  \\

        \bottomrule
    \end{tabular}}
    \label{tab:res_kernels_gen}
\end{table}

Further, we evaluate the models regularized using paired images from the \textit{Paired-public} dataset. The dataset contains only FC07/FC51 and FC07/FC55 kernel pairs. Besides the previous setup, this data does not contain COVID-19 lesions. Thus, we demonstrate that some methods depend on the semantic content and poorly generalize to kernel styles. The results are shown in Tab. \ref{tab:res_public}.

We highlight two main findings from these results. Firstly, consistency of the methods that operates with the decoder layers decreases to the level of Baseline; see the \textit{DANN (Dec)}, \textit{P-Cons}, and \textit{F-Cons (Dec)} rows in Tab. \ref{tab:res_public}. Our intuition here is that the decoder version of models can be more easily enforced to output the trivial predictions than the encoder one. Simultaneously, the images without COVID-19 lesions induce trivial predictions. Therefore, it might be easier for these models to differ the paired dataset from the source dataset by the semantic content and fail to align the stylistic features. Finally, we observe our method, \textit{F-Cons (Enc)}, to outperform the other adaptation methods training only on the publicly available data.

\begin{table}[h]
    \centering
    \caption{Comparison of all adaptation methods from Tab. \ref{tab:res_final} except \textit{FBPAug} trained on the \textit{Public-paired} dataset. All results are Dice Scores in the format \textit{mean} $\pm$ \textit{std} calculated from $5$-fold cross-validation. We highlight the consistency scores near or below \textit{Baseline} level in \textit{italic}. The best consistency scores are highlighted in \textbf{bold}.}
    %\vspace{.2cm}
    \resizebox{\textwidth}{!}{%
    \begin{tabular}{l c c c c c c c}
        \toprule
        
        & \multirow{2}{*}[-0.2em]{\shortstack{\textit{COVID-train}}} & \multirow{2}{*}[-0.2em]{\textit{COVID-test}}& \multicolumn{5}{c}{\textit{Paired-private} consistency} \\
        \cmidrule(lr){4-8}
         &  & & FC07/55 & FC07/51 & LUNG/SOFT & LUNG/STAND   & Mean\\
        
        \midrule
        Baseline & $.60 \pm .04$ & $.56 \pm .03$ & $.52 \pm .06$ & $.39 \pm .07$ & $.58 \pm .08$ & $.28 \pm .05$ & $.46 \pm .05$\\
        
        \cmidrule{1-8}
         
        DANN (Dec) & $.58 \pm .04$ & $.63 \pm .04$ & $.62 \pm .03$ & $.49 \pm .07$ & $\mathit{.60 \pm .03}$ & $\mathit{.30 \pm .04}$ & $\mathit{.53 \pm .03}$ \\
        
        \cmidrule{1-8}
        
        % DANN & $.58 \pm .05$ & $.64 \pm .02$ & $.84 \pm .02$ & $.70 \pm .02$ & $.86 \pm .03$ & $.66 \pm .06$ & $.78 \pm .02$ \\
        
        % \cmidrule{1-8}
         
        DANN (Enc) &  $.60 \pm .03$ &  $.64 \pm .02$ & $.75 \pm .02$ & $\mathbf{.64 \pm .05}$ & $.67 \pm .03$ & $.50 \pm .05$  & $.66 \pm .02$ \\
        
        \cmidrule{1-8}
        
        % P-Cons & $.59 \pm .04$ & $.61 \pm .01$ & $.65 \pm .05$ & $.60 \pm .02$ & $.77 \pm .01$ & $.47 \pm .04$ & $.63 \pm .03$\\
        
        % \cmidrule{1-8}
        
        P-Cons & $.53 \pm .03$ & $.58 \pm .03$ & $\mathit{.54 \pm .05}$ & $\mathit{.44 \pm .03}$ & $\mathit{.57 \pm .04}$  & $\mathit{.28 \pm .06}$ & $\mathit{.47 \pm .03}$  \\
        
        \cmidrule{1-8}
        
        F-Cons (Dec) & $.60 \pm .03$ & $.59 \pm .00$ & $\mathit{.54 \pm .05}$ & $\mathit{.47 \pm .05}$ & $\mathit{.64 \pm .05}$ & $\mathit{.31 \pm .06}$ &  $\mathit{.50 \pm .04}$ \\
        
        \cmidrule{1-8}
        
        % F-Cons &  $.57 \pm .03$ & $.64 \pm .03$ & $.88 \pm .01$ & $.72 \pm .04$ & $.83 \pm .02$ & $.70 \pm .05$ & $.80 \pm .01$ \\
        
        % \cmidrule{1-8}
        
        F-Cons (Enc) & $.59 \pm .04$ & $.64 \pm .02$ & $\mathbf{.80 \pm .02}$ & $\mathbf{.63 \pm .04}$ & $\mathbf{.71 \pm .02}$ & $\mathbf{.55 \pm .05}$ & $\mathbf{.70 \pm .02}$  \\
        
        \bottomrule
    \end{tabular}}
    \label{tab:res_public}
\end{table}

\subsection{Trade-off between consistency and segmentation quality}
\label{ssec:res:trade-off}

\begin{figure}[h]
    \centering
    \caption{Trade-off between the segmentation quality and consistency scores induced by the regularization parameter $\alpha$ (Sec. \ref{ssec:method:f-consistency}). The blue line corresponds to Dice Scores calculated on the \textit{COVID-train} dataset. The orange line corresponds to the consistency scores calculated on the \textit{Paired-private} dataset. The shaded areas correspond to the standard deviation along the Y-axis.}
    \label{fig:trade-off}
    \includegraphics[width=\textwidth]{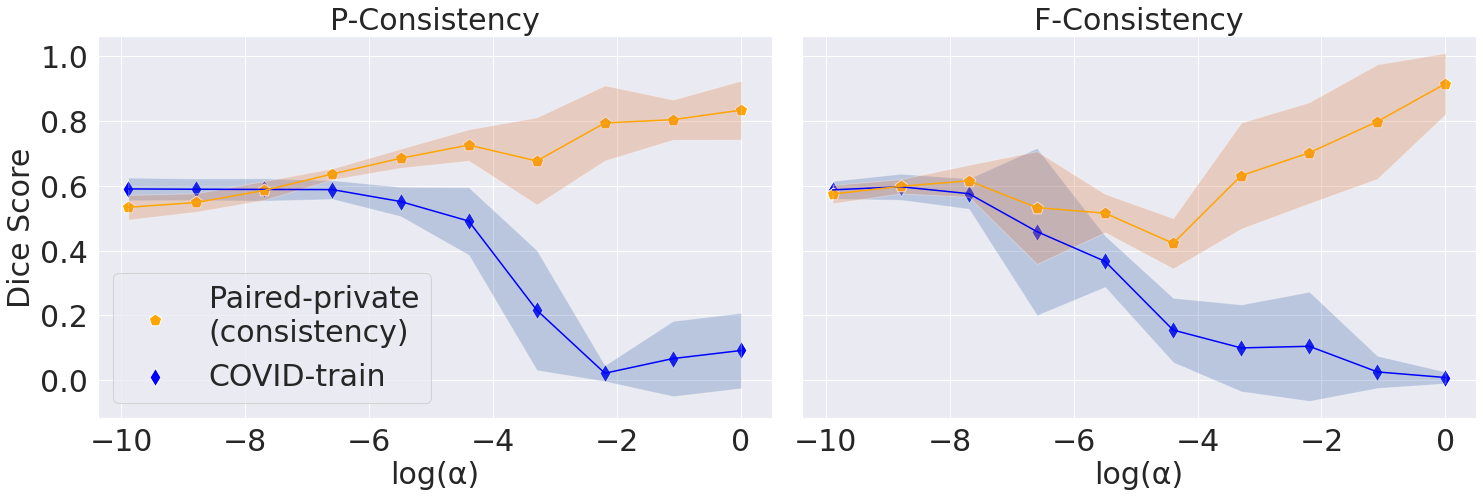}
\end{figure}

The main problem with maximizing the consistency score is converging to the trivial solution (empty predictions). Following the intuition from the previous section (\ref{ssec:res:generalization}), the problem is more acute for the adaptation methods that operate with the decoder layers. Thus, we vary $\alpha$, the parameter that balances the consistency and segmentation losses, for \textit{P-Consistency} and \textit{F-Consistency (Dec)}; see Fig. \ref{fig:trade-off}. The resulting trade-off follows the expected trend: consistency increases to $1$ and target Dice Score decreases to $0$ with increasing $\alpha$.

We use Dice Score on the \textit{COVID-train} dataset as a perceptual criterion to choose $\alpha$. We stop at the largest $\alpha$ value before Dice Score starts to drop: $10^{-3}$ for \textit{P-Cons}, $10^{-4}$ for \textit{F-Cons (Dec)} and $1$ for \textit{F-Cons (Enc)}. Here, we also ensure no overfitting under \textit{COVID-test} by using the cross-validation scores. However, we use \textit{Paired-private}, which participates in the final comparison, to calculate the consistency score. Firstly, we argue that using the \textit{Paired-public} dataset in this setup is incorrect. Paired-public does not contain COVID-19 lesions; thus, we can only measure the consistency of false positive predictions. Secondly, we choose $\alpha$ without considering consistency scores. Therefore, we also do not overfit under the consistency scores.

For the \textit{DANN} method, we choose $\lambda = 10^{-2}$ based on the best score on \textit{COVID-train}. Far from the optimal $\lambda$ values, \textit{DANN}'s prediction scores have the large standard deviation, so the trade-off cannot be observed. We also note that the adversarial approach does not explicitly enforce the trivial predictions. Hence, we report the trade-off only for the \textit{F-Consistency} and \textit{P-Consistency} methods.

\section{Discussion}
\label{sec:discussion}

Below, we summarize our results, discuss the most important limitations of our study, and suggest the possible directions for future work.

We have shown that the proposed \textit{F-consistency} significantly improves the performance on the \textit{target} domain compared to the \textit{baseline} model. However, we do not train the \textit{oracle} model, which indicates the upper bound for other methods in a domain adaptation task. The \textit{oracle} model should be trained via cross-validation on the \textit{target} domain. In our case, the \textit{target} domain contains only 9 images, which leads either to lower results due to the small size of the training set or high dispersion of the results. Therefore, we compare the adaptation methods only with the \textit{baseline} model and between each other.

In Sec. \ref{ssec:res:final}, our model achieves the highest results in terms of the consistency score. Contrary, the authors of \citep{orbes2019multi} observe a tendency of models to converge to trivial solutions using consistency loss. They assume that the models learn to distinguish domains for which they are penalized; thus, the models yield trivial but perfectly consistent predictions. Although we run the same setup with \citep{orbes2019multi}, we do not observe trivial predictions for our method. The latter is demonstrated through the whole Sec. \ref{sec:results}. Our intuition here is that the inner structure of domains and the semantic content of images are more diverse, preventing the model from overfitting under a specific domain.

One may argue that we could use the \textit{Paired-public} dataset (Sec. \ref{sssec:data:paired:public}) to calculate the consistency scores in our experiments since it also contains the paired images. Here, we highlight that the \textit{Paired-public} data was collected before the COVID-19 pandemic; thus, it does not contain COVID-19 lesions. Consequently, calculating the consistency scores on \textit{Paired-public}, we measure the consistency of false-positive predictions, i.e., noise. Therefore, we compare the models using the \textit{Paired-private} dataset (Sec. \ref{sssec:data:paired:private}) instead.

We highlight that adversarial and consistency-based methods depend on a diverse unlabeled pool of data; see Sec. \ref{ssec:res:generalization}. On the other hand, \textit{FBPAug} does not require additional data since it augments the images from \textit{source} dataset. One could think of this method as enforcing the consistency between the original and augmented image predictions using ground truth as a proxy. Simultaneously, we show that the models operating with the earlier layers to perform better. Therefore, we could train F-Consistency on the pairs of original and augmented with FBPAug images to achieve even better results. We leave the latter idea for future work.

% We also observe a large gap of consistency scores between setups described in Sec. \ref{ssec:res:final} and Sec. \ref{ssec:res:generalization}. These setups differs in both the number of training pairs and diversity of used reconstruction kernels. For all unsupervised methods we observe a decrease in consistency scores even for domains presented in both setups during training stage.  This gap is especially noticeable for consistency-based methods. Such observation indicates the need for a sufficient and diverse collection of paired images for the proposed method to work properly.

\subsection{Conclusion}
\label{ssec:discussion:conclusion}

We have proposed an unsupervised domain adaptation method, F-Consistency, to address the difference in CT reconstruction kernels. Our method uses a set of unlabeled CT image pairs and enforces the similarity between feature maps of paired images. We have shown F-Consistency outperformed the other adaptation and augmentation approaches in the COVID-19 segmentation task. Finally, through extensive evaluation, we have shown our method to better generalize on the unseen reconstruction kernels and without the specific semantic content.

\paragraph{Acknowledgments} The work was supported by the Russian Science Foundation grant 20-71-10134.% Computational experiments were powered by Zhores, a super computer at Skolkovo Institute of Science and Technology \citep{zacharov2019zhores}.

% \bibliography{main,triage,covid_algos,medical}
\bibliography{main,datasets}

\end{document}